\documentclass[12pt,epsfig,citesort,svgnames,dvipsnames]{article}
\usepackage{multicol}
\usepackage[all]{xy}
\usepackage{graphicx}
\usepackage{todonotes}
\usepackage{tikz-cd}
\usepackage{subfigure}
\usepackage{amsmath}
\usepackage{amssymb}
\usepackage{amsfonts}
\usepackage{amsthm}

\usepackage{amscd}
\usepackage{latexsym}
\usepackage[official]{eurosym}
\usepackage[english]{babel}
\usepackage[T1]{fontenc}
\usepackage[utf8]{inputenc}
\usepackage{authblk}
\usepackage{natbib}
\usepackage{xcolor}
\usepackage{hyperref}
\usepackage[nottoc]{tocbibind}
\hypersetup{
colorlinks=true,
citecolor=WildStrawberry,
anchorcolor=Black,
filecolor=cyan,
menucolor=magenta,
runcolor=cyan,
linkcolor=RoyalBlue,
linktoc=page, 
urlcolor=TealBlue}

\title{\bf {General Relativity and Background Independence}}
\author{\normalsize Antonio Vassallo}
\affil{\normalsize \emph{Warsaw University of Technology, Faculty of Administration and Social Sciences, Plac Politechniki 1, 00-661 Warsaw, Poland}\\antonio.vassallo@pw.edu.pl}
\date{}
\begin{document}

\maketitle

\begin{center}
\small
Preprint version. The final published version is available online at:
\url{https://doi.org/10.1016/B978-0-443-40622-5.00019-0}.\\
Forthcoming in S. O. Hansson (ed.), \emph{Comprehensive Philosophy of Science}, Elsevier.
\end{center}
\vspace{1em}

\begin{abstract}
General Relativity (GR) is widely regarded as a paradigmatic example of a background independent theory, a feature often taken to mark a decisive conceptual advance over its Newtonian and special relativistic predecessors. Yet the notion of background independence admits multiple formalizations, and its precise physical and philosophical significance remains contested. This chapter offers a systematic analysis of the strategies that have been proposed to capture background independence in classical spacetime physics. The discussion then turns to a central open question: whether, and in what sense, a successor theory of GR---such as a theory of quantum gravity---should be expected to inherit GR's background independence. Drawing on contemporary debates and a range of case studies, the chapter argues that background independence is best understood as a diagnostic and comparative tool rather than as a necessary physical requirement. The resulting perspective highlights both the conceptual virtues and the interpretive costs of eliminating background structures, and helps to explain why background independence remains an open problem in contemporary foundational research.
\end{abstract}

\newpage

\tableofcontents

\section{Introduction}\label{sec1}

General Relativity (GR) is routinely described as a \emph{background independent} theory. The slogan is ubiquitous in contemporary discussions of spacetime physics and its foundations, where it is often taken to capture one of Einstein's deepest insights into the nature of spacetime. The rough intuition behind the claim is familiar: unlike its Newtonian and special relativistic predecessors, GR does not presuppose a fixed spacetime stage on which physical processes unfold. Instead, spacetime geometry itself participates in the dynamics, responding to the distribution of matter and energy while simultaneously constraining physical motion. This picture has proven enormously influential, shaping both the interpretation of relativistic physics and the aspirations of contemporary research programs in quantum gravity.

Historically, the scientific relevance of the notion of background independence has strong roots in Einstein's own reflections on the discovery of GR. Indeed, Einstein regarded the transition from special relativity to GR as involving a radical reconceptualization of spacetime structure. This shift was motivated by the need to generalize the principle of relativity to all possible reference frames, in order to make the dynamical laws truly invariant under any coordinate transformation. Furthermore, Einstein wanted to implement the key realization---traceable back at least to Ernst Mach---that inertial structure itself demanded a dynamical explanation. In this respect, the contemporary debate inherits a set of conceptual concerns that first took shape with the emergence of classical mechanics and the philosophical reflections it prompted on the nature of space and time.

From its very beginnings, the formulation of a theory of motion raised questions about the status of the spatiotemporal structures invoked to articulate dynamical laws and to interpret their solutions. In this sense, the notion of a background can be approached through the role played by fixed spatiotemporal structures in grounding and constraining the dynamics, rather than through a precise formal criterion. Newtonian mechanics provides a natural starting point. In the Scholium to the \emph{Philosophiae Naturalis Principia Mathematica} \citep[][pp.~77--82]{25}, Newton treats space and time as entities that exist independently of material bodies and events. Spatial regions and temporal intervals possess their own identity, regardless of whether they are occupied or marked by physical processes. Newton is then a \emph{substantivalist} about space and time, regarding them as sui generis elements of the ontology presupposed by the theory.

This substantival commitment is accompanied by a stronger thesis. Newtonian space and time are absolute. They provide fixed standards for motion that do not depend on the distribution or state of matter. Absolute time ``flows equably,'' and absolute space remains ``immovable,'' serving as the reference structure with respect to which true motion and true acceleration are defined. These structures play an indispensable role in Newtonian dynamics. They make it possible to distinguish inertial motion from acceleration and to formulate the laws of motion in a coherent and universally applicable way. While Newton allows that God might have chosen different laws of motion, there is no suggestion that He could have placed bodies within a fundamentally different spatiotemporal framework. In modern terms, Newtonian space and time are fixed across all physically possible scenarios described by the theory. From this perspective, Newtonian space and time exemplify a paradigmatic case of background structures. Their features are fixed once and for all, they constrain the space of admissible dynamical histories, and they do not respond to the evolution of matter. At this stage, background structure and absolute structure appear to coincide. A fixed spatiotemporal scaffolding is required to make sense of the dynamics, and its invariance across all admissible situations marks it as part of the background.

Leibniz' relational critique brings sustained pressure to bear on this picture. In his correspondence with Samuel Clarke \citep{717}, Leibniz challenges the metaphysical coherence of absolute space and time by appealing to symmetry considerations and principles of sufficient reason. If the entire material universe were translated uniformly, or set into uniform motion as a whole, Newtonian mechanics would describe the resulting situation as distinct from the original one, even though no empirically observable difference would accompany the change. From Leibniz' perspective, absolute spatial positions and absolute states of rest encode distinctions without physical significance. They multiply physically possible scenarios without explanatory gain. Thus, \emph{relationalism} proposes an alternative organization of spatiotemporal facts. Spatial and temporal relations are grounded in relations among material bodies and events, and only those relations that make a difference to observable phenomena are retained in the ontology. The appeal of this stance lies in its economy. It avoids commitment to structures that perform no independent explanatory work. Yet this economy comes at a cost. By dispensing with fixed spatiotemporal standards altogether, relationalism struggles to ground certain inertial effects. Newton's bucket experiment provides the canonical illustration. The concave surface of the rotating water resists explanation solely in terms of the relative motion between the water and the bucket. Some further structure appears to be required to distinguish inertial from non-inertial motion. From this point of view, Leibniz' proposal faces a difficulty complementary to Newton's: where Newton postulates more structure than seems physically warranted, Leibniz postulates too little to support the full content of the dynamics. This tension, carefully analyzed in the modern literature, already reveals that background structure is not easily eliminated without loss \citep[see][for an informative articulation of the substantivalism--relationalism debate in pre-relativistic physics]{483a}.

The development of classical field theories and special relativity transforms this debate without resolving it. In this context, the introduction of Minkowski spacetime eliminates certain absolute structures while retaining others. Minkowski spacetime comes equipped with a fixed metric that determines its causal and inertial structures. Spacetime trajectories with non-zero proper acceleration remain objectively distinguished by their geometric attributes (i.e., the curve's intrinsic slope), and the spacetime geometry does not respond to the presence or evolution of matter fields. As a result, special relativistic theories continue to presuppose background structure, even though it differs in character from its Newtonian predecessor \citep[see, e.g.,][Ch.~III, \S~D4]{19}. One reason background structures have attracted sustained philosophical attention is that they limit a theory's explanatory reach: if certain spatiotemporal features are fixed rather than dynamical, the theory cannot explain why the world instantiates those particular structures rather than different ones. GR entered the scene against this backdrop and reshaped our understanding of spacetime by treating background independence as an essential, albeit elusive, feature of the theory's formulation.

The aim of what follows is to clarify how background independence in GR has been understood and formalized in subsequent debates, and to assess which of these approaches illuminate what is distinctive about general relativistic spacetime. The discussion is not intended to settle the substantivalism--relationalism dispute, to supply a necessary and sufficient definition of background independence, or to impose a single constraint on the form of a future theory of quantum gravity. Its positive contribution lies instead in articulating a disciplined way of comparing theories by tracking how their spatiotemporal structures enter into the space of physical possibilities they define. Framed in this way, background independence becomes a diagnostic tool for understanding what a theory takes to be physically real, and for evaluating how closely a candidate successor to GR preserves---or revises---the core insight that made spacetime dynamical in the first place.

\section{General Covariance and Background Independence}\label{sec3}

Einstein's development of GR was guided by a cluster of closely connected physical and methodological considerations, among which the status of spacetime structure played a central role (see \citealp{603}, for a comprehensive historical reconstruction of Einstein's path to GR based on an extensive analysis of original sources). Dissatisfaction with the fixed background geometry of special relativity, together with the ambition to incorporate gravitation into a relativistic framework, led Einstein to reconsider the role assigned to spacetime geometry in the formulation of physical laws. In this setting, Einstein was led to question whether the sharp separation between geometry and dynamics was physically well motivated. Gravitational phenomena suggested that spacetime geometry itself should participate in the dynamics, rather than merely providing a fixed arena for material processes. Indeed, the empirical indistinguishability (at least, locally) between inertial and gravitational effects pointed toward a conception of gravitation as a manifestation of spacetime structure itself. 

Pursuing this idea required a theoretical framework in which spacetime geometry could vary from one physically possible situation to another and could be determined, at least in part, by the distribution of matter and energy. Einstein's field equations realize this ambition by coupling spacetime geometry to the stress--energy tensor, thereby treating the metric as a dynamical field whose behavior depends on material sources. In this respect, GR departs decisively from both Newtonian mechanics and special relativity by allowing inertial structure itself to be sensitive to the physical state of the universe. In a nutshell, GR describes a reciprocal relation in which spacetime geometry constrains the motion of matter, while matter in turn shapes spacetime geometry \citep[][p.~5]{27}.

Disregarding the cosmological constant, which plays no role in the present discussion, and using natural units $G=c=1$, Einstein's equations can be written in intrinsic form as
\begin{equation}\label{efe3.1}
\mathbf{G}[\mathbf{g}]=8\pi\,\mathbf{T}[\mathbf{g},\phi].
\end{equation}

Here $\mathbf{g}$ is a Lorentzian metric field defined on a four-dimensional differentiable manifold $M$, i.e., a set of ``events'' whose local structure is sufficiently like that of a vector space to make the tools of calculus applicable. In general, $(M,\mathbf{g})$ cannot be treated globally as a single vector space because spacetime curvature implies locally varying standards of length and direction. $\mathbf{G}[\mathbf{g}]$ is the corresponding Einstein tensor, and $\mathbf{T}[\mathbf{g},\phi]$ is the stress--energy tensor associated with matter fields $\phi$, whose definition in general depends on $\mathbf{g}$ as well.  

A solution of (\ref{efe3.1}) is a triple $\langle M,\mathbf{g},\mathbf{T}\rangle$ satisfying the equation, interpreted as a physically possible spacetime scenario according to the theory. The space of such solutions constitutes the \emph{model space} of GR.

Equations of the form (\ref{efe3.1}) are written intrinsically in terms of tensor fields on $M$. For local calculations it is often convenient to introduce coordinates $\{x^{\mu}\}$ on an open region $U\subseteq M$ and express the equations in component form,
\begin{equation}\label{efe3.2}
G_{\mu\nu}=8\pi T_{\mu\nu},
\end{equation}

where the indices run from $0$ to $3$.

The passage from (\ref{efe3.1}) to (\ref{efe3.2}) amounts to a change of representation rather than a change of physical content. Given two coordinate systems $\{x^{\mu}\}$ and $\{\tilde{x}^{\mu}\}$ defined on $U$, with associated bases $\{e^{\mu}\}$ and $\{\tilde e^{\mu}\}$ at a point $P\in M$, the metric tensor can be written as
\begin{equation}\label{metric3.1}
\mathbf{g}=g_{\mu\nu}|_{P}\,e^{\mu}\otimes e^{\nu}
           =\tilde g_{\mu\nu}|_{P}\,\tilde e^{\mu}\otimes \tilde e^{\nu},
\end{equation}
where the repeated indices (one up and one down) are implicitly summed over. The components of the metric tensor in the two bases are related by the standard tensorial transformation law
\begin{equation}\label{metric3.2}
\tilde g_{\mu\nu}
=\frac{\partial x^{\rho}}{\partial \tilde x^{\mu}}
 \frac{\partial x^{\sigma}}{\partial \tilde x^{\nu}}
 g_{\rho\sigma}.
\end{equation}
Analogous relations hold for all tensorial quantities appearing in the field equations. As a result, the same intrinsic equations can be recovered from their component expressions in any coordinate system.

This representational freedom can be expressed more precisely using diffeomorphisms. A \emph{diffeomorphism} of a differentiable manifold $M$ is a smooth, bijective map $f:M\rightarrow M$ whose inverse is also smooth. The set of all such maps forms a group under composition, denoted by $\mathrm{Diff}(M)$. Any diffeomorphism $f\in \mathrm{Diff}(M)$ induces a natural transformation of the geometrical objects defined on $M$. For tensor fields, this transformation is implemented by the pullback operation. If $\boldsymbol{\Theta}$ is a tensor field on $M$, then $f^{*}\boldsymbol{\Theta}$ is the tensor field obtained by composing $\boldsymbol{\Theta}$ with $f$, together with the appropriate action on its indices. The pullback preserves tensorial relations and carries geometrical structure along the map in a consistent way. This intrinsic picture is closely related to the more familiar notion of a change of coordinates. Given a coordinate transformation $\{x^{\mu}\}\rightarrow\{\tilde{x}^{\mu}\}$ on a neighborhood of $M$, one can always associate a diffeomorphism $f$ such that, for each point $P$ in the domain, $x^{\mu}(f(P))=\tilde{x}^{\mu}(P)$. Conversely, given a diffeomorphism $f$ and a coordinate system $\{x^{\mu}\}$, one can define a new coordinate system $\{\tilde{x}^{\mu}\}$ by setting $\tilde{x}^{\mu}(P)=x^{\mu}(f(P))$. For this reason, invariance under arbitrary coordinate transformations can equivalently be expressed as invariance under the action of diffeomorphisms.

GR's descriptive flexibility can thus be expressed as a closure condition: for every diffeomorphism $f\in \mathrm{Diff}(M)$ and every model $\mathfrak{M}=\langle M,\mathbf{g},\mathbf{T}\rangle$, the pulled-back model $f^{*}\mathfrak{M}$ is also an element of the solution space of GR. This property captures what it means for GR to be a \emph{generally covariant} theory. In this sense, general covariance requires that the space of solutions be invariant under the action of the diffeomorphism group of the underlying manifold. Given the intrinsic formulation of the equations of GR, such invariance is expected: diffeomorphisms map geometrical objects to geometrical objects and solutions to solutions.

Einstein initially regarded general covariance as expressing a substantive physical insight, closely connected to the extension of the principle of relativity beyond the frameworks of Newtonian mechanics and special relativity. In both of those theories, the laws of motion take the same form in all inertial reference frames, and this restricted invariance under coordinate transformations is naturally interpreted as signaling the physical equivalence of those frames. At the same time, the restriction to inertial frames reflects the presence of a privileged inertial structure that grounds this equivalence and distinguishes inertial motion from acceleration. Einstein's ambition in developing GR was to remove this residual appeal to privileged inertial structure. He was therefore led to the idea that the laws of physics should retain their form under arbitrary coordinate transformations, rather than merely under transformations between inertial frames. On this reading, general covariance appeared to function as a generalized principle of relativity. The demand that the fundamental equations be invariant in form under arbitrary coordinate transformations was therefore taken to express a deep physical claim about the relational character of spacetime structure and about the absence of absolute inertial standards. In particular, Einstein associated general covariance with the view that spacetime points do not possess physical individuality independently of the fields defined on them. Physical content, on this conception, resides in coincidences of field values rather than in relations to a fixed spatiotemporal background. This interpretation is intelligible in light of the historical context from which GR emerged. Einstein understood the principle of relativity as a methodological guide rather than as a merely kinematical restriction. The progression from Galilean to Lorentz invariance already suggested that enlarging the covariance group of a theory could reflect a genuine deepening of its physical content. It was therefore natural for Einstein to regard the further enlargement to arbitrary coordinate transformations as marking a corresponding advance \citep[see][for an illuminating historical discussion of Einstein's development of GR, especially in relation to the notion of general covariance]{365}. 

This reading of general covariance was challenged almost immediately. In a paper published in 1917, Erich Kretschmann argued that the physical significance Einstein attributed to general covariance rested on a confusion between the mathematical form of a theory and its substantive dynamical content \citep{59}. His central claim was that general covariance, understood as invariance under arbitrary coordinate transformations, is a purely formal property that can be achieved by any spacetime theory, independently of whether that theory contains absolute structures or accords a privileged status to certain states of motion. Kretschmann's objection thus targets the inference from unrestricted covariance to the physical equivalence of all reference frames. Even if the equations of a theory retain their form under arbitrary coordinate transformations, this fact alone does not entail that the theory eliminates privileged inertial structure.

This point can be illustrated by a simple example. Consider a theory of a massless Klein--Gordon field $\phi$ propagating on a spacetime endowed with a fixed Minkowski metric $\boldsymbol{\eta}$. In inertial coordinates, the field equation takes the familiar form
\begin{equation}\label{kg0}
\eta^{\mu\nu}\phi_{,\mu\nu}=0,
\end{equation}
where the comma in the index denotes ordinary partial differentiation with respect to the corresponding coordinate. As written, (\ref{kg0}) is not generally covariant, since it holds only in coordinate systems adapted to the inertial structure determined by $\boldsymbol{\eta}$. One might therefore be tempted to regard the lack of general covariance as reflecting the presence of preferred inertial frames. However, this conclusion is premature. The equation can be rewritten in a manifestly generally covariant form by introducing a metric field $\mathbf{g}$ and the unique covariant derivative operator $\boldsymbol{\nabla}$ compatible with it (meaning $\boldsymbol{\nabla}\mathbf{g}=0$), with the understanding that $\mathbf{g}$ is everywhere flat and coincides with $\boldsymbol{\eta}$ up to diffeomorphism. The field equation then becomes
\begin{equation}\label{kg1}
g^{\mu\nu}\phi_{;\mu\nu}=0,
\end{equation}
the semicolon in the index representing covariant differentiation. Equation (\ref{kg1}) reduces to (\ref{kg0}) in inertial coordinate systems, but it retains the same form in any coordinate system whatsoever. In intrinsic notation, it can be written as
\begin{equation}\label{kg2}
\Box_{\mathbf{g}}\phi=0,
\end{equation}
where $\Box_{\mathbf{g}}$ denotes the d'Alembert operator associated with the metric $\mathbf{g}$. In this guise, the equation is formally invariant under arbitrary diffeomorphisms of the underlying manifold.

Despite its manifest general covariance, the theory described by (\ref{kg1})--(\ref{kg2}) remains a special relativistic theory in every physically relevant sense. The metric $\mathbf{g}$ is fixed across all solutions, does not respond to the behavior of the field $\phi$, and encodes a determinate inertial structure that distinguishes inertial from accelerated motion. The generally covariant reformulation leaves the theory's physical content unchanged. It has merely obscured, at the level of the equations' appearance, the presence of the background structure. The lesson of the example generalizes easily. Any spacetime theory, including Newtonian mechanics, can be rendered generally covariant by suitable reformulation. The availability of such reformulations shows that general covariance, taken by itself, does not impose substantive restrictions on the space of physically possible models of a theory. In particular, it does not guarantee the absence of fixed background structures, nor does it secure the physical equivalence of all reference frames. The Kretschmann objection therefore undermines the identification of general covariance with a generalized principle of relativity. 

This result has a clear and far-reaching consequence for the interpretation of GR. If general covariance can be achieved by any spacetime theory through suitable reformulation, then it cannot by itself account for what distinguishes GR from its predecessors. In particular, it cannot ground the claim that GR eliminates background structure in a substantive physical sense. The upshot is methodological as well as conceptual. Appeals to general covariance must be supplemented by a more fine-grained analysis of the theory's models and of the dependence relations encoded in them. The question of background independence cannot be settled by inspecting the covariance group of the equations alone. These considerations motivate the search for a notion of general covariance with genuine physical significance, a project that has generated an extensive literature in the philosophy of GR.

\section{Diffeomorphisms as Gauge Transformations: A Guide to Background Independence?}\label{subsec3.3}

The discussion so far has isolated a purely formal sense in which GR is generally covariant, and Kretschmann's point that this feature does not, by itself, distinguish GR from earlier spacetime theories. A natural next move aims at a sharper thesis: GR is distinctive because its covariance group $\mathrm{Diff}(M)$ functions as a gauge group. In \citet{369}, this proposal is formulated as a criterion of physically substantive general covariance and treated as a proxy for background independence, on the grounds that a spacetime theory cannot satisfy substantive general covariance if it contains background structures. The attraction of this proposal is clear. The relevant invariance now concerns the space of models and their physical interpretation, instead of the syntactic malleability of equations under coordinatization. Yet this strategy inherits a delicate burden. One must specify what counts as a gauge transformation in a way that does not trivialize the criterion and does not smuggle in, under the banner of ``interpretation,'' precisely the background structures that the criterion is meant to diagnose. 

A useful starting point is the notion of \emph{gauge freedom} in familiar field theories. In classical electromagnetism, one often introduces a scalar potential $\varphi$ and a vector potential $\mathbf{A}$ in terms of which the electric and magnetic fields are defined by
\begin{equation}\label{emfields-3.3}
\mathbf{E}=-\nabla \varphi-\frac{\partial \mathbf{A}}{\partial t},\qquad \mathbf{B}=\nabla\times \mathbf{A}.
\end{equation}
The fields $\mathbf{E}$ and $\mathbf{B}$ enter directly into the Lorentz force law and guide the motion of charged matter. The potentials $(\varphi,\mathbf{A})$ are not uniquely determined by $(\mathbf{E},\mathbf{B})$. For any sufficiently smooth scalar function $\lambda(\mathbf{x},t)$, if $(\varphi,\mathbf{A})$ is a potential pair, then so is $(\varphi',\mathbf{A}')$ defined by
\begin{equation}\label{gaugetransf-3.3}
\varphi'=\varphi-\frac{\partial \lambda}{\partial t},\qquad \mathbf{A}'=\mathbf{A}+\nabla \lambda.
\end{equation}
A short calculation using (\ref{emfields-3.3}) verifies that $\mathbf{E}'=\mathbf{E}$ and $\mathbf{B}'=\mathbf{B}$. The transformation (\ref{gaugetransf-3.3}) therefore connects distinct mathematical representations of the same electromagnetic field configuration. This example supplies a general template. A \emph{gauge group} acts on field configurations so that gauge-related configurations represent the same physical possibility; physical quantities are accordingly invariant under the group action, or at least well defined on the space of gauge-equivalence classes. The decisive point is interpretive rather than merely formal: whether a symmetry counts as gauge depends on how representational redundancy is identified and how physical situations are individuated across mathematical presentations. This is exactly where GR becomes contentious. 

In the modern literature, a common proposal says that GR is substantively generally covariant because $\mathrm{Diff}(M)$ relates gauge-equivalent models. In \citet{174} the proposal has two parts: (i) the solution space is diffeomorphism invariant, and (ii) for any $d\in \mathrm{Diff}(M)$ the models $\mathfrak{M}$ and $d^{*}\mathfrak{M}$ represent the same physical possibility. Clause (ii) carries the interpretive weight. It effectively generalizes the lesson commonly drawn from the hole argument in GR \citep[see][for a thorough introduction to the debate surrounding this argument]{12}: treating diffeomorphically related models as distinct physical possibilities yields unpalatable underdetermination, while treating them as representational variants restores determinism and better reflects the practice of extracting physical content from generally relativistic solutions. \citet[][p.~445]{174} emphasizes that the substantive requirement should be respected as a ``datum'' supplied by a widely shared view in physics, while remaining open to philosophical scrutiny.

It is useful to see concretely how treating $\mathrm{Diff}(M)$ as gauge reshapes observables. In many field theories one speaks of field values ``at a point,'' e.g., the electromagnetic field at $(t,x,y,z)$. In GR this idiom lacks invariant content: a diffeomorphism $f\in \mathrm{Diff}(M)$ sends $P$ to $f(P)$ while dragging metric and matter fields along, and $P$ carries no fixed identity under such transformations. Quantities like $g_{\mu\nu}(P)$ or $\phi(P)$ therefore fail, by themselves, to define gauge-invariant observables. What survives are correlations among fields. If a model contains a scalar clock field $T$ and another scalar $\phi$, one can form the relational quantity ``the value of $\phi$ when $T=\tau$.'' Under a diffeomorphism both fields are dragged, while the correlation is preserved: the manifold point where $T=\tau$ may change, yet the value of $\phi$ at that condition does not. Diffeomorphism invariant observables thus take the form of correlations among dynamical fields. The point is not merely technical: localization, duration, and distance are recovered from dynamical relations rather than read off from a fixed background grid, so questions about observables and about background independence intertwine. This way of understanding observables in GR has been forcefully articulated by \citet{56}, who argues that background independence manifests itself most directly in the theory's conception of what is physically measurable. Rickles emphasizes that background independence should be tracked down to the theory's empirical content.\footnote{The technical and philosophical debate about what counts as observable in GR is huge and has obvious ramifications for the quest for a theory of quantum gravity. The interested reader can refer to \citet{67} for a clear entry point into the relational-observables strategy.}

The contrast with background dependent theories helps to clarify what is at stake. Consider again the Klein--Gordon theory on Minkowski spacetime introduced in \S~\ref{sec3}. One may write the dynamics in generally covariant form as $\Box_{\mathbf{g}}\phi=0$, but the theory carries an additional constraint that fixes $\mathbf{g}$ to be flat and Minkowskian. In such a theory, the metric plays the role of a background structure: it is the same in all models (up to trivial redescription), and it fixes a non-trivial symmetry group, namely the isometry group of Minkowski spacetime, called the Poincar\'e group $\mathrm{Iso}(\boldsymbol{\eta})$. The transformations in this group are isometries of the Minkowski metric in the sense that, for any $d \in \mathrm{Iso}(\boldsymbol{\eta})$, $d^{*}\eta_{\mu\nu}=\eta_{\mu\nu}$. The Poincar\'e group has a clear physical significance, namely, it underwrites conservation laws and the classification of inertial motions. Diffeomorphisms in $\mathrm{Diff}(M)$ act far more broadly than Poincar\'e transformations. A generic diffeomorphism $f$ will not preserve the Minkowski metric; it will map $\eta_{\mu\nu}$ to a distinct metric field $f^{*}\eta_{\mu\nu}\neq\eta_{\mu\nu}$. If $\eta_{\mu\nu}$ is treated as representing a fixed physical structure, then such a diffeomorphism changes that structure and therefore changes the physical situation as represented by the model. In this sense, diffeomorphisms fail to be symmetries in the relevant sense in background dependent theories: they do not preserve the structures that the interpretation takes to be physically fixed, and they therefore cannot be treated as mere redescription. The contrast mirrors the electromagnetic case. Gauge transformations change potentials while preserving $\mathbf{E}$ and $\mathbf{B}$. In a background dependent spacetime theory, a generic diffeomorphism changes the metric that fixes inertial and causal structure. When that metric is taken as fixed and physically significant, only its isometries count as genuine symmetries. A fixed background therefore separates $\mathrm{Diff}(M)$, understood as a covariance group, from its symmetry or invariance group, which preserves the physically significant structure.

The GR case is often presented as different because the metric is not fixed across models. This invites the idea that no non-trivial subgroup of $\mathrm{Diff}(M)$ is selected by a fixed spacetime structure, and that diffeomorphisms can therefore be treated as gauge transformations relating equivalent representations. This is the point at which substantive general covariance  comes into view. Once the proposal is stated that way, the guiding thought becomes: background independence tracks the absence of fixed geometric structures that survive across all models as physically significant objects. If all diffeomorphisms are gauge, then manifold points do not carry physical identity independently of the fields; the metric field can then be understood as dynamical geometry rather than fixed stage-setting. This is the methodological payoff: it promises a criterion that appears to separate GR from special relativistic field theories on Minkowski spacetime.

A prominent contemporary criticism of this line of thought is developed by \citet{370}. His central complaint targets the claim that diffeomorphism invariance (plus the gauge reading of $\mathrm{Diff}(M)$) provides a route to background independence. The core point is simple to state: background dependent theories admit diffeomorphism invariant formulations, including formulations that look, at first glance, like GR's. This complicates any attempt to read off background structure directly from the mere presence of a diffeomorphism group treated as a symmetry of the equations, or of the model class. A major source of confusion, on Pooley's diagnosis, concerns what spacetime coordinates are doing in background dependent theories. Coordinates are sometimes treated as pure gauge even when they are tacitly tied to background geometry in the way they represent inertial structure. Once that conceptual role is made explicit, it becomes less plausible that diffeomorphism invariance, as a bare formal property, reliably separates background dependent from background-independent theories \citep[][\S~10]{791}. Pooley's conclusion is that diffeomorphism invariance cannot serve, on its own, as an informative criterion of background independence, because one can manufacture diffeomorphism invariant reformulations of theories whose background dependence remains intact. 

As an example, due to \citet[][pp.~110--112]{47}, consider again the massless Klein--Gordon equation. One may begin with the familiar special relativistic formulation on Minkowski spacetime. One can then write a generally covariant version by introducing a general Lorentzian metric $g$ and adjoining the constraint $\mathbf{Riem[g]}=0$, meaning that the Riemann curvature tensor is identically null, so that only flat metrics are permitted. In that dynamicized formulation, the equations look GR-like: all fields appear to live on the manifold, obey their own dynamical equations, and the system is closed under diffeomorphisms. \citet{800} sharpen Pooley's moral: diffeomorphism invariance is a well-defined formal property, while background independence admits multiple, interpretation-dependent readings. For that reason, diffeomorphism invariance can constrain the form of a theory without settling, on its own, whether the theory contains background structure. Thus, if the idea of substantive general covariance is to retain physical bite, a principled way is needed to distinguish between cases in which diffeomorphism invariance reflects a genuine dynamical symmetry of the theory and cases in which it arises merely from rewriting the equations in covariant form. A prima facie plausible way to draw this distinction is to ask whether diffeomorphism invariance is built into the action principle that generates the equations of motion rather than being imposed after the fact at the level of the field equations.

\section{Background Independence and Action Principles}

An \emph{action principle} specifies a theory by means of an action functional $S$ defined on a space of field configurations. For a set of fields $\Psi$ on a spacetime manifold $M$, the action typically takes the form
\begin{equation}\label{action}
S[\Psi]=\int_{M}\mathcal{L}(\Psi,\partial\Psi)\,,
\end{equation}
where $\mathcal{L}$ is a Lagrangian density constructed from the fields and their derivatives. The equations of motion are obtained by requiring that $S$ be stationary under arbitrary variations of the fields with suitable boundary conditions. A transformation of the fields is said to be a \emph{variational symmetry} if it leaves the action invariant (up to a boundary term), and hence maps solutions of the Euler--Lagrange equations to solutions.

In many familiar cases, variational symmetries have a direct physical interpretation. Global symmetries of the action are associated, via Noether's first theorem, with conserved quantities. Local symmetries, by contrast, often signal gauge freedom: they indicate representational redundancy, in the sense that distinct field configurations related by the symmetry can correspond to the same physical state of affairs. From this perspective, identifying the gauge group of a theory amounts to identifying the group of local variational symmetries of its action.

In the case of GR, this line of thought appears especially compelling. The Einstein--Hilbert action,
\begin{equation}\label{EH}
S_{\mathrm{EH}}[g_{\mu\nu}]=\frac{1}{16\pi}\int_{M} R[g_{\mu\nu}]\sqrt{-\det|g|}\,d^{4}x,
\end{equation}
(where $R$ is the Ricci scalar and $\det|g|$ denotes the determinant of the matrix representing $\mathbf{g}$ in a given coordinate system) is invariant under arbitrary diffeomorphisms of the spacetime manifold. Therefore, this invariance goes beyond the field equations derived from the action, being a property of the variational principle itself. Moreover, diffeomorphism invariance of the action is associated, via Noether identities and the Bianchi identities, with the covariant conservation of the stress--energy tensor on shell, a feature naturally connected to the dynamical character of spacetime geometry. These considerations motivate the claim that diffeomorphisms in GR play the role of genuine gauge transformations, in closer analogy with gauge symmetries in Yang--Mills theories. Seen in this light, the appeal to action principles promises to strengthen the notion of substantive general covariance. 

The key question is whether action-based diffeomorphism invariance excludes background dependent theories that merely imitate GR's formal profile, such as the diffeomorphism-invariant reformulations discussed by Pooley and Giulini. This is exactly where \citet[][]{366}'s construction becomes decisive. Sorkin shows how to cast a flat-space scalar field theory into a diffeomorphism invariant form derived from an action by introducing an auxiliary Lagrange multiplier field $\lambda_{\mu\nu\rho\sigma}$ that enforces flatness. Sorkin emphasizes that, in this formulation, conservation of stress--energy emerges as a formal consequence of diffeomorphism invariance, closely mirroring the GR case. \citet[][pp.~115--116]{47} presents the same idea and stresses that the $\lambda$-field can enlarge the solution space in a way that seems to add more than mere redescription, unless one is careful about what counts as physical. \citet[][\S~8]{791} uses Sorkin's example to argue that even an action-based criterion that identifies background independence with $\mathrm{Diff}(M)$ being a variational symmetry group will overgenerate: it risks classifying a reformulated special relativistic theory as background independent. The example forces a methodological choice about the status of the auxiliary variables. If they represent additional physical degrees of freedom, the reformulated theory is not equivalent to the original background dependent one. If they are treated as pure surplus, the reformulation becomes an elaborate redescription that should not affect verdicts about background dependence. Either way, formal diffeomorphism invariance plus an ad hoc decision to ignore auxiliary structure cannot settle the issue. One needs principled criteria for when extra fields carry physical content and when they function as representational fluff.

One influential attempt to refine the action-based strategy is due to \citet{792}, who argue that the shortcomings exposed by Sorkin-style constructions arise from an overly permissive identification of gauge symmetries with diffeomorphism invariance at the level of the action. According to their view, not all variational symmetries should be treated on a par. What matters for physical content is the distinction between trivial gauge transformations, which generate no observable charges, and non-trivial symmetries, which are associated with conserved quantities or boundary charges via Noether's theorems. Freidel and Teh propose to link background independence to the structure of Noether charges. The guiding idea is that non-dynamical structures support non-vanishing symmetry charges, while genuine gauge freedom does not. On this proposal, diffeomorphisms that act as pure gauge transformations in GR fail to generate physical charges because there is no fixed background geometry relative to which such charges could be defined. By contrast, in reformulated special relativistic theories of the Sorkin type, the presence of auxiliary fields and hidden absolute structure allows for the recovery of non-trivial charges, revealing that the apparent diffeomorphism invariance does not reflect genuine background independence. \citet{793} has challenged this strategy by arguing that the proposed refinement does not ultimately succeed in blocking Kretschmann-style trivialization. Struyve's analysis shows that one can construct diffeomorphism invariant actions for background dependent theories in which the would-be gauge symmetries fail to generate non-trivial charges, even though absolute structure remains present in the theory. From this perspective, the absence of Noether charges does not reliably track the absence of background structure. The Freidel–Teh proposal and Struyve's reply sharpen the dialectic without settling it. The upshot is that action principles and variational symmetries illuminate important aspects of gravitational dynamics, while no purely variational criterion straightforwardly isolates what is distinctive about GR's treatment of spacetime structure.

\section{Absolute Structure and the Anderson--Friedman Proposal}\label{sec4}

\citet{47} sharpens the diagnosis of covariance-based criteria by disentangling two notions that are frequently conflated: covariance and invariance. In Giulini's framework, equations of motion are written schematically as $F[\gamma,\Phi,\Sigma]=0$, where $\Phi$ denotes dynamical fields and $\Sigma$ collects non-dynamical structures. Covariance under a group $G\subseteq \mathrm{Diff}(M)$ permits $\Sigma$ to transform alongside the dynamical variables; invariance requires that transformations map solutions to solutions while leaving $\Sigma$ fixed. General covariance, therefore, guarantees a geometrically well-defined formulation on a manifold while remaining silent about whether a theory contains background structure: fixed elements may persist inside $\Sigma$ without obstructing covariance. This matters because the substantive general covariance proposal---read as the claim that $\mathrm{Diff}(M)$ is gauge---aims to upgrade covariance into an invariance-plus-interpretation thesis; doing so requires that structures treated as fixed either be eliminated or incorporated into the dynamical sector in a non-trivial way. Giulini's diagnosis thus redirects attention away from covariance considerations alone and toward the theory's inventory of absolute structures, which may remain concealed unless the formalism explicitly isolates them as separate fields.

The most influential attempt to provide a direct criterion for diagnosing fixed spacetime structure is due to James Anderson and Michael Friedman. Anderson's guiding intuition is that background structure is marked by a failure of reciprocity: some geometrical elements constrain the behavior of matter without themselves being affected in return. He formalizes this idea through the notion of an \emph{absolute object}. Roughly speaking, an object is absolute if it has the same geometrical form in all models of the theory, up to the action of the theory's covariance group \citep[][\S~4.3]{66}. Friedman's refinement of Anderson's proposal embeds the notion of absolute structure within a more explicit account of theoretical equivalence \citep[][Ch.~II, \S~2]{15}. The key move is the introduction of a notion of \emph{d-equivalence} between models, intended to capture physical sameness modulo diffeomorphism freedom. Two models $\langle M,\Phi_{1},\dots,\Phi_{n}\rangle$ and $\langle M,\Psi_{1},\dots,\Psi_{n}\rangle$ of a spacetime theory $T$ are said to be d-equivalent if, for every point $P\in M$, there exist neighborhoods $A$ and $B$ of $P$ and a diffeomorphism $f:A\to B$ such that, on the overlap $A\cap B$, each field $\Psi_i$ is obtained as the pullback of $\Phi_i$ by $f$. D-equivalence thus requires local diffeomorphic matching of field values around each point, rather than the existence of a single global diffeomorphism relating entire models. This locality is essential: it allows Friedman to abstract from global topological or boundary differences while still capturing the idea of representational redundancy rooted in spacetime diffeomorphisms. On this basis, Friedman defines an absolute object as follows. A geometrical object $\Phi_i$ is absolute in a spacetime theory $T$ if and only if, for any two models $\langle M,\Phi_{1},\dots,\Phi_{n}\rangle$ and $\langle M,\Psi_{1},\dots,\Psi_{n}\rangle$ of $T$, the fields $\Phi_i$ and $\Psi_i$ are d-equivalent. Intuitively, an absolute object is one whose local form is fixed across all physically admissible models of the theory, up to diffeomorphism, so that it does not encode any model-specific physical information. 

Within this framework, absolute objects play a structural role in fixing the symmetry group of the theory. Following Anderson, the symmetry group is defined as the largest subgroup of the covariance group that leaves all absolute objects invariant. The fewer absolute objects a theory contains, the larger its symmetry group. In special relativistic field theories, the covariance group is $\mathrm{Diff}(M)$, but the symmetry group collapses to the Poincaré group $\mathrm{Iso}(\boldsymbol{\eta})$ of isometries that preserve the Minkowski metric. In GR, by contrast, given that no absolute objects are present, the symmetry group coincides with the full diffeomorphism group. This connection between absolute structure and symmetry provides a precise sense in which GR differs from its predecessors: not merely in admitting a diffeomorphism invariant formulation, but in failing to admit any formulation that is not diffeomorphism invariant \citep[][\S~6]{791}.

Despite its virtues, the framework faces several well-known difficulties. First, some features of GR, such as spacetime dimensionality or the Lorentzian signature of the metric, are fixed across all models in a trivial sense: they are insensitive to diffeomorphism transformations (or changes of coordinates). These are what \citet[][\S~II]{353} call \emph{confined variables}, and what \citet[][\S~2]{354} terms \emph{confined objects}. Yet these are not geometric objects of the theory but constraints on the class of admissible models, and therefore fall outside the category of absolute objects as defined by Anderson and Friedman \citep[cf.][p.~354]{354}. The presence of confined objects thus highlights a limitation of the framework. Consider, for instance, the Lorentzian signature of the metric. As \citet{811} emphasizes, the Lorentzian local structure of spacetime in GR implements the assumption that special relativity holds locally; in that sense, it carries genuine physical significance even though it is trivially fixed across models. One might therefore be tempted to regard it as a kind of absolute structure. Whether this intuition should be endorsed---on pain of making the notion of background structure too coarse---remains a matter of debate. The point, however, is that this important foundational question cannot even be formulated within the Anderson--Friedman framework. This leads \citet[][p.~270]{812} to conclude that characterizing spacetime theories in terms of their invariance properties is not misguided in itself, but that the identification of absolute structures ultimately rests on physically motivated interpretive judgments.

A second challenge is developed by \citet{354}. Pitts shows that the presence or absence of absolute objects can depend sensitively on the choice of variables and geometrical language. Reformulations that replace fixed structures with dynamically constrained fields---for example via Lagrange multipliers---can eliminate absolute objects without altering empirical content. If taken at face value, this would render background independence an artifact of representation. Pitts, therefore, proposes a refinement inspired by Anderson's own methodological strictures: absolute objects should be identified only among geometrical entities that do indispensable explanatory work throughout the theory's models. This refinement excludes both globally and locally irrelevant variables, and blocks standard counterexamples such as the Jones--Geroch dust \citep[][\S~5]{354}.

A related worry arises from Torretti's constant-curvature example \citep{795}. Torretti considers theories in which each model has constant curvature, though the value varies from model to model. Intuitively, such theories appear to presuppose significant background structure, yet the metric itself fails to qualify as absolute on Anderson's or Friedman's definitions. Pitts's response shows that this conclusion is premature: while the metric is not absolute, the conformal metric density is invariant across all models. Once tensor densities and non-standard geometrical objects are admitted, the framework succeeds in identifying a non-trivial background element after all \citep[][\S~6]{354}.

A further complication for Anderson's appeal to the action–reaction principle emerges from historical and conceptual analyses of Einstein's own reasoning. \citet{438} argue that the idea that Newtonian spacetime violates action–reaction rests on a retrospective reinterpretation rather than on Newton's own conception of space. Newton denied that absolute space exerts forces or plays a causal role in the production of inertia; if so, its failure to be acted upon constitutes no violation of reciprocity. Conversely, in GR, the claim that the metric field both acts on and is acted upon by matter is less straightforward than often assumed \citep{582}. In this respect, the action–reaction principle does not provide a stable metaphysical marker for distinguishing background from dynamical structure, even in the paradigmatic cases that originally motivated it.

Taken together, these results clarify both the power and the limits of the Anderson--Friedman approach. The notion of absolute structure provides a sharp diagnostic for analyzing spacetime theories and their symmetry groups, especially when refined to exclude irrelevant variables. At the same time, its sensitivity to formulation and interpretation suggests that background independence is better understood as a graded and comparative notion, tracking how much structure is fixed and how tightly it constrains the space of physical possibilities.

\section{Background Independence in Degrees}\label{sec5}

As the previous discussion makes clear, GR does not eliminate all fixed structure, since certain features remain invariant across all solutions. \citet{311} takes this situation as a point of departure, proposing that GR should be understood as weakening, rather than abolishing, background structure. On this view, background independence comes in degrees, determined by how much geometrical and dynamical information is fixed in advance rather than solved for by the field equations.

To see how Belot's verdicts are generated, it is crucial to understand how his proposal distinguishes geometrical from physical degrees of freedom. A spacetime theory, Belot argues, must be equipped with a \emph{geometrization}: a rule that associates to each solution a spacetime geometry, together with a criterion of geometrical equivalence, typically diffeomorphism equivalence. Relative to such a choice, the geometrical degrees of freedom are the parameters required to distinguish inequivalent spacetime geometries across the theory's solutions. Physical degrees of freedom are defined by first quotienting the space of solutions by gauge equivalence, understood as the identification of solutions that the theory itself forces us to regard as physically the same in order to avoid indeterminism. The remaining parameters label genuinely distinct physical possibilities. Background dependence is then assessed by comparing these two counts. In fully background dependent theories, every solution is assigned the same geometry, so there are no geometrical degrees of freedom at all. In fully background independent theories, every physical degree of freedom corresponds to a geometrical one, so two solutions represent the same geometry if and only if they are gauge equivalent. Intermediate cases arise whenever some, but not all, physical degrees of freedom register as variations in spacetime geometry. Belot's program therefore measures background independence by tracking how tightly changes in the theory's physical state are tied to changes in its geometrical structure.

A central feature of Belot's proposal is its explicit dependence on interpretation. Whether two models count as representing distinct physical possibilities depends on how one interprets the variables appearing in the formalism. Hence, Belot's account generalizes a lesson already implicit in earlier debates over general covariance, gauge freedom, and absolute objects: background independence concerns what a theory says about the world, not merely how it is mathematically formulated.

The graded character of background independence becomes especially clear in theories that occupy intermediate positions. Belot discusses examples in which geometry is partially constrained across models, or where only certain aspects of spacetime structure are dynamical. In such cases, it is neither accurate nor illuminating to classify the theory as simply background dependent or independent. Instead, the theory exhibits background independence with respect to some degrees of freedom while retaining fixed structure in others. This perspective allows one to accommodate cases that resist sharp classification under Anderson--Friedman-style criteria, without collapsing into triviality.

The framework's interpretive dependence is illustrated in Belot’s examples, where the same formalism admits distinct physical readings, yielding correspondingly different verdicts about background independence. A particularly illuminating case involves a theory describing the non-linear propagation of a scalar field over Minkowski spacetime.

Consider a theory specified by the following equations:
\begin{subequations}\label{chenex2}
\begin{equation}\label{minchioski2}
\mathbf{Riem}[\mathbf{g}]=0,
\end{equation}
\begin{equation}\label{mesorotto2}
\boldsymbol{\Box}_{\mathbf{g}}\phi=-4\pi\phi^{3}\mathbf{T},
\end{equation}
\end{subequations}
where $\mathbf{g}$ is a Lorentzian metric, $\phi$ a scalar field, and $\mathbf{T}$ a source term representing the stress–energy distribution. Equation (\ref{minchioski2}) constrains $\mathbf{g}$ to be flat, while (\ref{mesorotto2}) governs the dynamics of $\phi$ relative to that geometry. Interpreted in the natural way, this theory describes a scalar field propagating on Minkowski spacetime. The metric $\mathbf{g}$ is fixed across all models of the theory, even though it is formally presented as a solution of a field equation. On this interpretation, the theory is clearly background dependent: physically distinct solutions correspond to different configurations of $\phi$ relative to the same fixed spacetime geometry.

Now consider a metric $\mathbf{g'}$ defined by:
\begin{subequations}\label{nor}
\begin{equation}\label{nor1}
\mathbf{g'}=\phi^{2}\mathbf{g},
\end{equation}
\begin{equation}\label{nor2}
\phi=(-\det|g'|)^{\frac{1}{8}},
\end{equation}
\begin{equation}\label{nor3}
\mathbf{g}=\mathbf{g'}(-\det|g'|)^{\frac{1}{4}}.
\end{equation}
\end{subequations}
These relations allow one to trade the pair $(\mathbf{g},\phi)$ for a single geometrical object $\mathbf{g'}$, provided $\mathbf{g'}$ is conformally flat---meaning that, in suitable coordinates, it differs from the flat metric only by a position-dependent scale factor (as shown in Eq.~\eqref{nor1}), so angles are preserved while lengths are locally rescaled.

When the theory (\ref{chenex2}) is rewritten in terms of $\mathbf{g'}$, one obtains the equivalent system:
\begin{subequations}\label{chenex3}
\begin{equation}\label{minchioski3}
\mathbf{Weyl}[\mathbf{g'}]=0,
\end{equation}
\begin{equation}\label{mesorotto3}
\mathbf{R}[\mathbf{g'}]=24\pi\mathbf{T}.
\end{equation}
\end{subequations}
Here, $\mathbf{Weyl}[\mathbf{g'}]$ denotes the Weyl curvature, capturing the conformal, shape-distorting degrees of freedom of spacetime curvature, while $\mathbf{R}[\mathbf{g'}]$ encodes how matter locally sources curvature.


Equations (\ref{chenex3}) therefore describe a conformally flat metric---since the Weyl curvature vanishes---whose remaining curvature is entirely sourced by $\mathbf{T}$. On their face, they closely resemble a generally relativistic theory with a dynamical spacetime geometry. Indeed, if one considers two models $\mathfrak{M}_{1}=\langle M,\mathbf{g'}_{1},\mathbf{T}_{1}\rangle$ and $\mathfrak{M}_{2}=\langle M,\mathbf{g'}_{2},\mathbf{T}_{2}\rangle$, in general $\mathbf{g'}_{1}\neq \mathbf{g'}_{2}$ and $\mathbf{T}_{1}\neq \mathbf{T}_{2}$. Relative to this formulation, the theory appears to exhibit background independence: variation in the physical degrees of freedom is accompanied by variation in spacetime geometry.

At first sight, this situation is paradoxical. The systems (\ref{chenex2}) and (\ref{chenex3}), given the relations (\ref{nor}), are mathematically equivalent. Yet the former seems manifestly background dependent, while the latter seems background independent. Belot's resolution of this apparent paradox rests on the recognition that background independence is not a purely formal property of a set of equations, but a feature of an interpreted theory. If the theory is interpreted as describing a scalar field propagating on Minkowski spacetime, then $\mathbf{g}$ represents the genuine spacetime geometry and $\phi$ represents a physical field defined upon it. In this context, the composite object $\mathbf{g'}$ is merely a convenient way of packaging the pair $(\mathbf{g},\phi)$ into a single geometrical object. It does not represent an independent spacetime metric, and the apparent dynamical character of $\mathbf{g'}$ reflects nothing more than the dynamics of $\phi$ on a fixed background. If, by contrast, the theory is interpreted as describing a conformally flat spacetime metric $\mathbf{g'}$ coupled to matter, then the relations (\ref{nor}) are read in the opposite direction. The Minkowski metric $\mathbf{g}$ now plays the role of a fixed auxiliary structure introduced to facilitate the formulation of the theory, while $\phi$ functions as a surrogate encoding the conformal factor of the physical metric. On this interpretation, the theory describes a genuinely dynamical geometry and qualifies as background independent in Belot's sense. This example reinforces Belot's broader claim that background independence comes in degrees and cannot be identified with any single formal criterion. What matters is how strongly variation in geometry is tied to variation in physical content, relative to a chosen interpretation of the theory's variables.

\citet{794}'s analysis provides further support for this shift in perspective. While sympathetic to Anderson and Friedman's emphasis on absolute structure, Samaroo argues that their framework is best understood as highlighting a particular dimension along which background dependence can vary. From this standpoint, Belot's graded conception does not discard the insights of absolute-object diagnostics but situates them within a broader taxonomy. Absolute structures remain a salient indicator of background dependence, yet their presence does not exhaust the ways in which geometry may function as a background.

\citet{789} has urged caution about the broader significance often attributed to background independence. He argues that even a refined, interpretation-sensitive account does not support strong non-empirical claims about theory choice. The fact that a theory scores higher on some measure of background independence does not, by itself, provide reason to regard it as more likely to be true. This critique does not undermine Belot's framework, but it does clarify its proper role. The value of a graded conception lies in organizing and clarifying debates about spacetime structure, not in delivering decisive methodological verdicts.

\citet[][\S~3.6.4]{790} raises two connected worries about Belot's framework. The first concerns what \citet{381} has called the ``priority of geometry'': in relativistic theories, whether a given field should count as spacetime geometry is not fixed by the formalism alone but depends on how that field earns its chronogeometric significance. Belot's proposal presupposes an identification of the geometrical sector of a theory which is prior to dynamics, whereas approaches inspired by Anderson and Brown suggest that such identifications themselves require substantive physical argument. This opens the door to accounts of background independence that treat all fields on a par, rather than singling out geometry in advance. The second worry is more technical. Belot's criterion can deliver counterintuitive verdicts in cases where geometrical degrees of freedom vary across models without being dynamically responsive within each model, and in theories---such as source-free Einstein–Maxwell theory---where distinct matter configurations can correspond to the same spacetime geometry. In such cases, Belot's definitions either misclassify intuitively background dependent structures as background independent, or else imply that familiar relativistic theories possess more background structure than one might have expected. Read's diagnosis is that Belot's framework tracks only one aspect of background independence, leaving open a gap between dynamical degrees of freedom and what should count as genuine spacetime structure.

\section{Leibnizian--Machian Strategies: Eliminating Surplus Structure}\label{sec6}

Mach's critique of absolute space and time, most forcefully articulated in \citet{110}, targets the physical origin of inertial effects themselves. Its core demand is that appeals to unobservable spacetime structures be replaced by explanations grounded in the observable distribution and motion of matter. Einstein took this challenge seriously in developing GR, explicitly formulating a principle bearing Mach's name, i.e., the requirement that spacetime structure be determined by the material contents of the universe. Although GR realizes this ideal only partially, it sharpens the Machian problem by turning it into a precise dynamical question. The Leibnizian--Machian program radicalizes this insight by aiming to eliminate external spatiotemporal structure altogether, rather than merely rendering it dynamical.

The modern Leibnizian--Machian program takes its mathematically precise form in the work of Julian Barbour and Bruno Bertotti, most notably in \citet{84,83}. Their point of departure is explicitly Leibnizian: fundamental physical laws should not refer to absolute positions, absolute orientations, or an external universal clock. Only relations among physical systems are to be admitted as primitive. What makes their contribution decisive is that this demand is implemented as a structural requirement on the dynamical theory itself. A theory satisfies Mach's principle, in their sense, when its equations of motion are invariant under the full group of transformations that alter only the external spatiotemporal embedding of a system while preserving all internal relations. This group includes translations, rotations, and reparametrizations of time. If two descriptions differ only by such a transformation, they represent the same physical situation. A theory that treats them as distinct therefore contains surplus structure in exactly the sense targeted by Leibnizian critiques of absolute space and time.

The technical mechanism that implements this idea is the method of \emph{best matching}. One begins with the standard configuration space $Q$ of a system. For $N$ particles in Newtonian mechanics, $Q=\mathbb{R}^{3N}$. This space carries a natural action of a group $G$ representing spatiotemporal redundancies, typically the Euclidean group of translations and rotations and, in later implementations of the program (see, e.g., \citealp{419}), scale transformations. Thus, the physical configuration space is not $Q$ itself but the quotient $Q_{\mathrm{phys}}=Q/G$, whose points represent equivalence classes of configurations related by transformations that change only absolute position, orientation, or scale. What remains are purely relational degrees of freedom. When scale is also quotiented out, the resulting space is \emph{shape space}, whose points encode only the intrinsic form of a system, given by ratios of distances and relative angles. Dynamics is then defined as geodesic motion on this quotient space with respect to a metric induced by the best-matching procedure. Time is no longer an external parameter but an ordering of configurations along a curve in shape space. In this way, background independence is achieved by construction in a straightforward sense: space and time are eliminated from the fundamental description, rather than being rendered dynamical (irrespective of whether the starting particle theory is classical or quantum; see, e.g., \citealp{433}).

This relational strategy can be extended beyond particle mechanics. The central result of Shape Dynamics (SD) is that, under appropriate conditions, GR can be reformulated as a theory of evolving spatial conformal geometry rather than four-dimensional spacetime geometry \citep{528}. In this reformulation, Einstein gravity is equivalent to a theory invariant under local conformal transformations of three-dimensional spatial metrics. The physical degrees of freedom are encoded in trajectories in the space of conformal three-geometries, making SD a direct field-theoretic analogue of relational particle mechanics. The role played there by absolute position and orientation is played here by refoliation freedom and local scale. By quotienting out these structures, one obtains a theory whose fundamental arena is a relational configuration space of spatial shapes.

\citet{410} proposes to turn this construction into a general definition of background independence. On his account, a theory is background independent when its space of physical states is obtained by quotienting a larger configuration space by all transformations that correspond to redundant spatiotemporal structure. The criterion is therefore not whether certain geometrical objects are dynamical, but whether the theory has been formulated directly on the reduced space $Q/G$, where $G$ captures the full group of background symmetries. In this sense, GR in its standard formulation remains only partially background independent, since it retains a differentiable manifold as the necessary basis to represent spacetime. SD and related Machian theories go further by eliminating additional layers of spatiotemporal structure through explicit quotienting. This elimination of external structures is further implemented in more recent developments of the program, where relational evolution in shape space is made even more intrinsic by renouncing any external parametrization of the dynamical curve, so that change is represented directly as an ordering of shapes rather than as evolution with respect to an external parametrization \citep{746,732}.

The Machian program hence offers a clear and principled way to eliminate background structure. However, this ambition comes at a cost. The resulting theories are technically more involved than their spatiotemporal counterparts and, crucially, they are non-local. In the case of standard SD, for example, this means that the true generator of evolution (and some of the gauge conditions) is determined by solving global elliptic constraints---so the instantaneous state at one point depends on data over the entire spatial slice, not just its neighborhood. Concretely, the conformal factor and the global Hamiltonian are fixed by solving equations of the Lichnerowicz–York type, making time evolution an intrinsically non-local functional of the three-geometry \citep[][]{544}. Even though later refinements of SD avoid this technique, still the theory frames the evolution of the shape variables in terms of functionals of the full 3-geometry, rather than by purely point-wise local evolution equations \citep{780}. This tension is not accidental. It mirrors the original Newton--Leibniz dispute in modern form. Newton's framework provided powerful and tractable tools for doing physics, at the price of heavy metaphysical commitments. Leibniz's relationalism offered ontological economy, at the price of conceptual and technical complexity. The contemporary Machian program inherits this legacy. It shows that background independence can be realized, at least in principle, in a precise technical sense. It also shows why such a realization is neither straightforward nor cost free.

\section{Should a Successor Theory Inherit GR's Background Independence?}\label{sec7}

A central hope of many quantum gravity programs is to find a successor theory to GR: a framework that reproduces GR in an appropriate classical limit while also providing a consistent quantum description of gravitational phenomena. A natural question then arises. Should such a successor theory inherit GR's background independence? In light of the above discussion, the question is delicate because background independence is not a single, settled criterion. The present section therefore treats background independence as a disputed inheritance claim: what, if anything, about GR's way of eliminating fixed spatiotemporal structure should be carried over into whatever comes next?

\citet{60}'s influential case for background independence is best read as an attempt to elevate a broadly Einsteinian lesson into a methodological constraint on quantum gravity. The core idea is conditional: if GR is empirically superior to its Newtonian and special relativistic predecessors partly because it does not presuppose a fixed spacetime arena, then a quantum successor that reinstates such an arena would amount to a step backward rather than a genuine continuation. Smolin frames this as the modern expression of the relational–absolute debate, and maintains that a viable quantum theory of gravity must preserve GR's dynamical treatment of geometry and its associated symmetry structure. This demand is tied to a stronger explanatory ambition: a final theory should not leave unexplained fixed structures or parameters, since doing so would undermine its claim to fundamental status.

Two familiar criticisms target this inheritance claim. Read's objection targets the elevation of background independence from a heuristic guide to a rationalist constraint on acceptable quantum gravity theories. One worry concerns scope: classical explications of background independence are often tailored to field theories on differentiable manifolds and may not carry over to genuinely non-classical kinematical frameworks. A second worry is methodological: even if one can articulate a surviving ``spirit'' of background independence, insisting on it as a necessity risks either imposing an unnecessary straightjacket or lapsing into principled vagueness. From an empiricist perspective, if a theory saves the phenomena while violating a favored explication of background independence, the fault lies with the explication rather than with the theory \citep[][pp.~130--131]{790}.

Pooley's criticism presses a related point in a simpler metaphysical register: granting that fixed geometry invites ``why this and not otherwise?'' questions, it does not follow that a successor theory must forbid brute geometric facts. A background geometry could simply be a brute element of the theory without being fixed by a priori principles in the way Smolin suggests \citep[][footnote 14]{791}. In short, Smolin's explanatory demand may be too strong to function as a constraint on theory construction rather than as a philosophical preference about what a final theory ought to explain.

De Haro's work helps to reorganize this dialectic by distinguishing different roles that background independence talk can play, especially in quantum gravity contexts shaped by gauge/gravity duality. He separates a \emph{minimalist} role---closely modeled on GR's own background independence and functioning as a baseline consistency requirement for gravitational theories---from an \emph{extended} role, according to which not only bulk dynamics but also boundary or initial structures should be dynamically determined. De Haro cautions against treating this extended notion as mandatory, while allowing that it may retain heuristic value in exploring generalizations of existing frameworks \citep[][\S~2.3.4]{796}. This distinction is introduced because many contemporary examples sit precisely at the boundary between ``dynamical bulk'' and ``fixed boundary data.'' In asymptotically AdS gravity, for example, specifying boundary conditions is part of what it is to pick out a solution; and those boundary conditions can be viewed, depending on the question at hand, either as necessary auxiliary structure or as a mark of residual background dependence. De Haro stresses that it is a misconception to think that one must pick some special ``AdS boundary conditions'' as though they were dictated by the theory; rather, (roughly speaking) arbitrary boundary data can be used to determine solutions, and the real interpretive question is what status to give the resulting boundary structure. This opens conceptual space for a successor theory that is background independent in the minimalist sense while still making essential use of boundary structure, without thereby collapsing into the background dependent paradigm of Newtonian theory or special relativity.

\citet[][\S~5.3.5]{790} raises a sharp technical-philosophical concern in this neighborhood by discussing De Haro's appeal to the diffeomorphism anomaly, i.e., his view that some diffeomorphisms relate physically inequivalent models \citep[][p.114]{796}. In a nutshell, this feature is often expressed in terms of ``small'' versus ``large'' diffeomorphisms. Small diffeomorphisms preserve the asymptotic or boundary conditions and generate gauge redundancy. Large diffeomorphisms change the asymptotic data and therefore map a solution to a physically distinct one. The worry, as Read formulates it, is that De Haro sometimes presents the anomaly as undermining the Kretschmann-style point that any theory can be given a generally covariant formulation; yet this risks conflating trivial general covariance with substantive diffeomorphism invariance. Read argues that even if an anomaly breaks diffeomorphism invariance, it does not follow that the theory cannot be rendered generally covariant in the trivial sense, since the coordinate-free language of differential geometry remains available. The correct moral, then, is not that the anomaly refutes Kretschmann, but that it forces care in distinguishing different notions of covariance and invariance, and in stating which notion is actually relevant to the form of background independence under discussion. De Haro's own reply strategy (in the framework Read is engaging) is to insist on precisely this kind of disambiguation, and to connect the physically relevant questions to which diffeomorphisms should count as gauge, especially once boundary conditions and ``large'' diffeomorphisms are in play \citep[][\S~2.3]{797}. The upshot for the inheritance question is that diffeomorphism invariance, boundary structure, and background independence come apart in ways that are invisible if one treats general covariance as a single yes/no affair. It is interesting to note that, in the string theory literature, the pressure to use boundary structure is part of what makes the duality empirically and conceptually tractable. This is one reason why, even independently of De Haro's specific proposals, authors such as \citet{435} have treated boundary conditions and asymptotic structures as philosophically central to what background independence can mean in the presence of holography, rather than as mere technical scaffolding to be removed at the end.

The possibility that a successor theory may need to relax rather than radicalize GR-like background independence becomes especially vivid once one focuses on observables, understood as more than just formal invariants in the abstract but as quantities intended to represent material structures and their measurable behavior. \citet{38}'s discussion of gravitational energy is instructive in this regard. The gravitational energy of a system in GR is notoriously subtle, and standard well-behaved notions of total energy typically require additional structure---most familiarly, asymptotic conditions and the associated asymptotic symmetries that pick out a meaningful notion of time translation at infinity. The conceptual moral is that even in paradigmatically generally covariant theories, physically significant quantities can depend on auxiliary background-like structures because without such structure the relevant quantity is ill-defined. In other words, certain kinds of background structure can function as conditions of definability for physically significant quantities, rather than as fixed arenas that merely host the dynamics. This point also resonates with the broader theme, emphasized above, that boundary conditions can be physically indispensable rather than merely conventional.

A closely related moral is developed by \citet{429} in the context of five-dimensional extensions of GR. The core idea is that one can formulate a natural ``GR-desideratum'' for successor theories: in the classical limit they should be diffeomorphism-gauge theories that do not postulate spatiotemporal structure beyond what is encoded in the metric on an appropriate semi-Riemannian manifold. The five-dimensional case study then shows how an overly strict pursuit of that desideratum can yield interpretive pathologies. In induced-matter style frameworks \citep{122}, for example, full $\mathrm{Diff}(M_{5})$ gauge freedom can make putatively physical four-dimensional matter quantities depend on gauge choice, threatening to demote them to gauge fluff. The proposed remedy is to complete the theory by adding further structure (for instance, a preferred foliation) that is fixed across models and hence counts as a background in a straightforward sense. The payoff is that the added structure can restore a stable notion of physical observable content. On this diagnosis, background structures can function as elements that complete a theory's physical content, preventing otherwise dynamical quantities from collapsing into gauge-dependent artifacts. From this perspective, it is not implausible that a successor to GR may need to reintroduce some background-like structure for reasons internal to interpretability and empirical coherence, even while remaining GR-like in other structural respects.

The morals are therefore mixed---and that is precisely the point. Smolin is right to insist that GR's dynamical treatment of geometry represents a hard-won lesson, and that reinstating a fixed spacetime arena would risk undoing what made GR empirically and conceptually distinctive. Read and Pooley are equally right to resist turning that lesson into a necessity claim about quantum gravity, given both the plurality of background independence notions and the legitimate roles that brute or auxiliary structure may play in physical theorizing. De Haro's minimalist–extended distinction offers a productive way forward: a GR-like absence of fixed bulk metric structure can be treated as a baseline consistency constraint, while stronger demands concerning boundary or initial data remain optional and context-sensitive. The unresolved task is to articulate, in a principled way, how much of GR's dynamical treatment of spacetime must be inherited, and how much fixed structure is required for a theory to support well-defined observables. The persistence of this tension reflects a genuine difficulty in theory construction, and may help to explain why a fully satisfactory successor to GR has yet to emerge.

\section{Conclusion}\label{sec8}

The analysis developed in this chapter supports a clear and non-trivial lesson. Background independence, as exemplified by GR, cannot be captured by a single formal criterion, nor can it be treated as an unqualified methodological requirement. Instead, it expresses a structural trade-off at the heart of spacetime theorizing. As emphasized by \citet{467}, nomically necessary and dynamically inert structures carry a substantial explanatory cost. When a theory postulates fixed spatiotemporal elements that do not participate in the dynamics, it relinquishes the ability to explain why physical processes unfold as they do, rather than otherwise. Such structures mark brute features of the world, immune to dynamical scrutiny. At the same time, those very features perform indispensable interpretive work. Fixed structures sharply define physical sameness across models, ground symmetry principles, and make central physical magnitudes---such as inertia and energy---unambiguous and tractable.

GR occupies a delicate intermediate position within this landscape. By dynamizing spacetime geometry, it secures a remarkable gain in explanatory depth concerning inertial and gravitational phenomena. Yet it does so at the price of conceptual complexity, forcing a rethinking of observables and symmetries. Subsequent attempts to push further toward full background independence amplify this tension: gains in explanatory ambition are often accompanied by losses in interpretive clarity and in the stable definition of physical quantities. The question of inheritance must therefore be framed accordingly. A successor theory to GR cannot simply maximize background independence without regard for its interpretive consequences, nor can it reintroduce fixed structure without sacrificing the explanatory lessons that GR made unavoidable. In this sense, ongoing debates over background independence register a genuine structural tension in fundamental physics. That tension is part of GR's enduring philosophical legacy---and part of the unfinished business facing any theory that aims to succeed it.

\pdfbookmark[1]{Acknowledgements}{acknowledgements}
\begin{center}
\textbf{Acknowledgements}:
\end{center}
I wish to thank the editor of the philosophy of physics section, Alastair Wilson, for his invitation to contribute to the collection. I also gratefully acknowledge financial support from the National Science Centre, Poland, grant No. 2023/49/B/HS1/01091.

\bibliography{biblio}

\end{document}